\documentclass[12pt]{article}
\usepackage{graphicx}

\def\napoli{Johannes Gutenberg, University of Mainz (DE)}
\def\napolidue{prencipe@slac.stanford.edu}
\def\napolitre{\small on behalf of the \babar Collaboration}

\def\Title#1{\begin{center} {\Large #1 } \end{center}}
\def\Author#1{\begin{center}{ \sc #1} \end{center}}
\def\Address#1{\begin{center}{ \it #1} \end{center}}

\newenvironment{Abstract}{\begin{quotation}  }{\end{quotation}}
\newenvironment{Presented}{\begin{quotation} \begin{center} 
             PRESENTED AT\end{center}\bigskip 
      \begin{center}\begin{large}}{\end{large}\end{center} \end{quotation}}

\def\babar{$BABAR~$}

\def\beq{\begin{equation}}
\def\eeq#1{\label{#1}\end{equation}}
\def\eeqn{\end{equation}}

\def\beqa{\begin{eqnarray}}
\def\eeqa#1{\label{#1}\end{eqnarray}}
\def\eeqan{\end{eqnarray}}

\let\bar=\overbar

\def\Dslash{\not{\hbox{\kern-4pt $D$}}}
\def\dslash{\not{\hbox{\kern-2pt $\del$}}}

\def\msb{{\bar{\ssstyle M \kern -1pt S}}}

\begin{document}
\begin{titlepage}

\vfill
\Title{Recent charmonium results from \babar}
\vfill
\Author{ELISABETTA PRENCIPE}
\Address{\napoli \\ \napolidue \\ \napolitre}
\vfill
\begin{Abstract}
Recent results in the field of spectroscopy from the ~\babar experiment are reported, with particular attention to the new states observed in ISR and $\gamma \gamma$ interactions using the full \babar data sample. We confirm the states Y(4260) and Y(4360) with higher precision and for the first time the state Y(4660) is observed. We do not confirm the Y(4008) state reported by Belle. In addition, the analysis of the invariant mass of the $J/\psi \omega$ system produced in $\gamma \gamma$ interactions is presented in confirmation of the Belle observation of the X(3915) in this process.  
\end{Abstract}
\vfill

 \begin{Presented}
The 5$^{th}$ International Workshop on Charm Physics\\ 
Honolulu, Hawai'i (USA),  May 14-17, 2012
\end{Presented}
\vfill
\end{titlepage}
\def\thefootnote{\fnsymbol{footnote}}
\setcounter{footnote}{0}

\section{Introduction}
$~~~$The past 10 years represent a very hightly productive period in the field of spectroscopy in physics, as many states not expected in theoretical predictions have been observed in different experiments, several decay modes and in different physics processes.  

Because of their mass values and their decays to charmonium states, they are believed to have hidden charm content, and since they do not fit into the predicted charmonium spectrum they are usually referred  to as {\it charmonium-like} states. The spin-parity assignment is in most cases unknown, therefore the standard naming convention cannot be used and they are named as X, Y or  Z. They have been extensively investigated as possible candidates for non-conventional mesons, tetraquarks, glueballs or hybrids~\cite{XYZrev}.

There are several ways to produce such X, Y, Z states at $B$ factories: 
\begin{itemize}
\item production in B decays: these are color-suppressed decays, typically processes with a $B$ meson decaying to $X$($c \bar c$)$K$;
\item two-photon production, for final $X$ states with the quantum number $J$ $\neq$ 1; the process involved is $e^+e^- \rightarrow e^+e^- c \bar c$;
\item  double charmonium production: this implies positive $C$ parity of the final chamonium $X$($c \bar c$) state;
\item production in continuum $e^+e^-$ interactions: vector $X$ initial states are produced in a process in which a photon is emitted from the $e^+e^-$ initial state, followed by the production of a $c \bar c$ state at lower $e^+e^-$ center-of-mass energy.
\end{itemize}

In this paper we summarize recent preliminary results from \babar in $\gamma \gamma$ interactions, and in production in $e^+e^-$ continuum via initial state radiation (ISR) processes.

\section{New results in ISR processes}
\subsection{Analysis of the $J/\psi \pi^+ \pi^-$ system produced via ISR}

$~~~$ The first result is related to the analysis of $e^+e^- \rightarrow \gamma_{ISR}J/\psi \pi^+ \pi^-$ at \babar\cite{vale}, where a clear resonant state is observed. The study of this state has been already shown from several experiments\cite{babar1,belle1,cleo1,cleoc1}. In the previous \babar analysis a clear state at 4260 MeV/c$^2$ was observed using a data sample corresponding to 211 fb$^{-1}$ integrated luminosity; now we confirm this state with higher precision using data corresponding to integrated luminosity 464 fb$^{-1}$. We do not confirm the resonant state reported by Belle  at 4008 Mev/c$^2$ in a similar analysis. The state Y(4260) is still a puzzling object; we know that its quantum numbers are $J^{PC}$ = 1$^{--}$, because it is produced in an ISR process, but further investigation of this state is necessary. 
\begin{figure}[htb]
\centering
\includegraphics[height=2.5in]{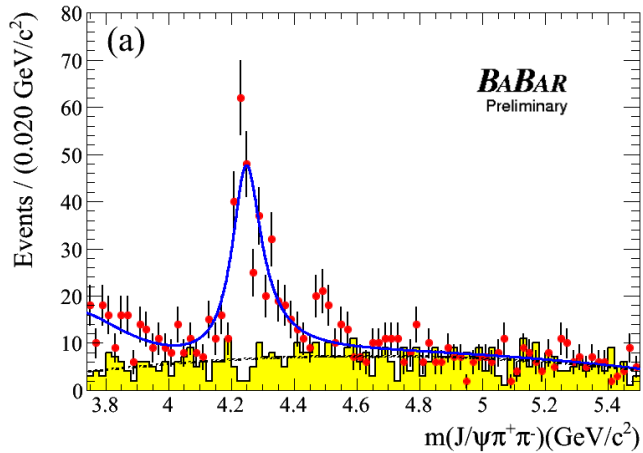}
\caption{Invariant mass of the $J/\psi \pi^+ \pi^-$ system from the process $e^+e^- \rightarrow \gamma_{ISR}J/\psi \pi^+ \pi^-$.}
\label{fig:Y4260}
\end{figure}
Fig.~\ref{fig:Y4260} shows the invariant $J/\psi \pi^+ \pi^-$  mass for the full \babar dataset.

An extended-maximum-likelihood fit is performed to the data from the $J/\psi$ signal region and simultaneously to the background distribution from the $J/\psi$  sidebands (yellow histogram) in the region [3.74$-$5.5] GeV/c$^2$.
The fit function incorporates the mass-dependence of efficiency and luminosity,
and uses a relativistic Breit-Wigner (BW) signal function for the Y(4260), a 3$^\mathrm{rd}$-order polynomial to describe the background, and an
empirical exponential function to describe the excess of events below 4 GeV/c$^2$, which may result from the $\psi(2S)$ tail and a possible $J/\psi \pi^+ \pi^-$ non-resonant contribution. Results for the mass and  width of the Y(4260) are reported in Table 1.

The di-pion mass distribution from the Y(4260) signal region ([4.15$-$4.45]
GeV/c$^2$) was investigated. We define $\theta_\pi$  as the angle between the $\pi^+$ direction and that of the recoil $J/\psi$, both in the di-pion rest frame.
The distribution of $cos \theta$, which must be
symmetric, is consistent with S-wave behaviour ($\chi^2$/NDF = 12.3/9, probability = 19.7$\%$). 
\begin{figure}[htb]
\centering
\includegraphics[height=2.5in]{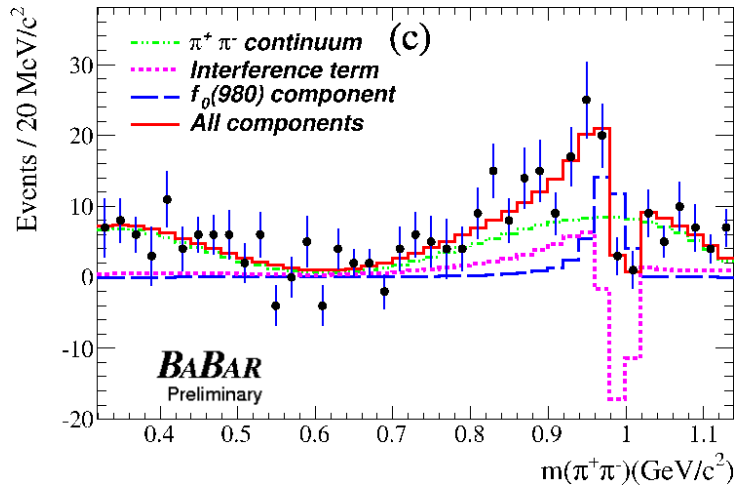}
\caption{Fit to the $\pi^+ \pi^-$ invariant mass distribution.}
\label{fig:dipion1}
\end{figure}

The $\pi^+ \pi^-$ mass distribution is shown in Fig.~\ref{fig:dipion1}, where a structure peaking around the mass of $f_0$(980) is observed. The displacement  of the peak toward lower mass suggest the possibility of interference between the $f_0$(980) and a $\pi^+ \pi^-$ continuum distribution.

A simple model is used to describe the di-pion mass distribution, namely the square
of an amplitude consisting of the coherent sum of a non-resonant component,
motivated by the QCD multipole expansion, and an $f_0$(980) amplitude; the relative strength and phase of these components are free to vary in the fit to the data. The mass dependence of the $f_0$(980) amplitude and phase is from the \babar analysis of Ref.\cite{dominance}.

The result of the fit is shown in Fig.~\ref{fig:dipion1}, and it is found that the Y(4260) decay to $J/\psi f_0$(980) is not a dominant contribution to the $J/\psi \pi^+ \pi^-$ final state (BR $\sim$17$\%$).

\subsection{Analysis of the $\psi$(2$S$) $ \pi^+ \pi^-$ system produced via ISR}

$~~~$ The first evidence for a resonant state at 4360 MeV/c$^2$ produced in  $e^+e^- \rightarrow \gamma_{ISR}\psi(2S)$ $\pi^+ \pi^-$ came from \babar\cite{babar2}; this was  then confirmed by Belle\cite{belle2}. We present an update of the \babar discovery, and the confirmation of the Y(4660), which was not observed in the previous \babar analysis because of the limited statistics. For this new analysis in \babar we made use of the full datasets collected at the $\Upsilon(nS)$, n=2,3,4. As these two new states, the Y(4360) and Y(4660), are seen in an ISR process, we conclude that their quantum numbers are the same as those of the virtual photon: $J^{PC}$= 1$^{--}$. 
\begin{figure}[htb]
\centering
\includegraphics[height=2.0in]{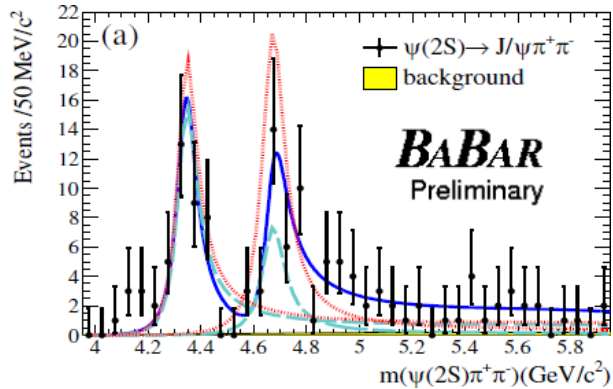}
\caption{Invariant mass of the $\psi$(2S) $ \pi^+ \pi^-$ system from the process $e^+e^- \rightarrow \gamma_{ISR} \psi $(2S) $\pi^+ \pi^-$.}
\label{fig:Y4660}
\end{figure}

The $\psi(2S) \pi^+ \pi^-$ invariant  mass spectrum,  corresponding to an integrated luminosity of 530 fb$^{-1}$, is shown in Fig.~\ref{fig:Y4660}. 

An unbinned extended-maximum-likelihood fit is performed to the  $\psi(2S) \pi^+ \pi^-$ invariant mass distribution for the events in the $\psi(2S)$ signal region, and simultaneously to the background distribution corresponding to the $\psi$(2S) sideband regions (Fig.~\ref{fig:Y4660}). In Table 1 the mass and width values obtained for the two resonances are reported.

\begin{figure}[htb]
\centering
\includegraphics[height=1.8in]{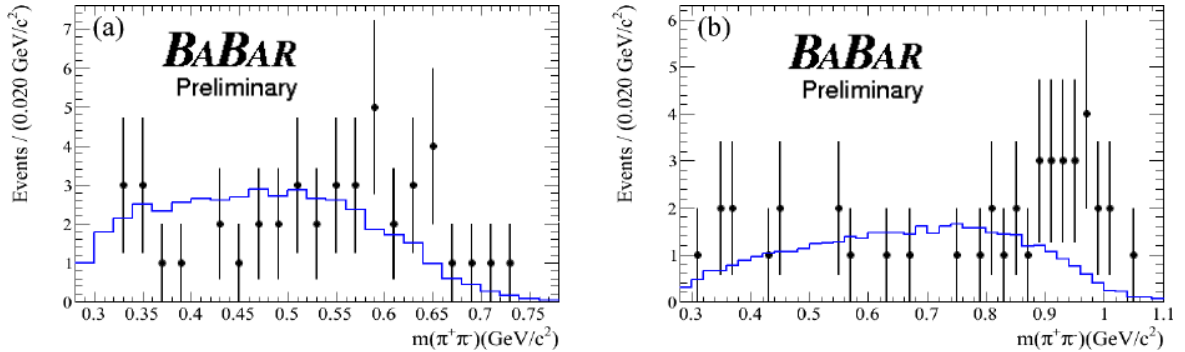}
\caption{Dipion invariant mass distribution for the signal region  of the (a) Y(4360) and (b) Y(4660).}
\label{fig:dipion2}
\end{figure}

The $\pi^+ \pi^-$ invariant mass spectrum was examined for each of the two resonance regions (Fig.~~\ref{fig:dipion2}). Unfortunately we cannot arrive at any clear conclusions  due to the low statistics available. However, the events tend to cluster around the $f_0$(980) region of $\pi^+ \pi^-$ invariant mass for the Y(4660) signal region (Fig.~\ref{fig:dipion2}(b)).

\begin{table}[!htb]
\caption{Mass and width values tor the states observed in the ISR processes (\babar preliminary).}
\begin{center}
\begin{tabular}{lrcccl}
\hline
\hline
Resonance & Mass (MeV/c$^2$)      & $\Gamma$ (MeV) \\ \hline
Y(4260)   & 4224 $\pm$ 5 $\pm$ 4  &  114$^{+16}_{-15}$ $\pm$ 7\\  
Y(4360)   & 4340 $\pm$ 16 $\pm$ 9 &  94 $\pm$ 32 $\pm$ 13 \\  
Y(4660)   & 4669 $\pm$ 21 $\pm$ 3 &  104 $\pm$ 48 $\pm$ 10\\ \hline 

\end{tabular}
\end{center}
\label{table1}
\end{table}

\section{New results in $\gamma \gamma$ interactions}
\subsection{Analysis $\gamma \gamma \rightarrow \eta_c \pi^+ \pi^-$}

 $~~~$ The process under study is the production of an unknown state X, where $\gamma \gamma \rightarrow X \rightarrow \eta_c \pi^+ \pi^-$. In this case $X$ is considered to be one of the following candidates: $\chi_{c2}$(1P), $\eta_c$(2S), X(3872) or $\chi_{c2}$(2P), and the $\eta_c$ decay process is  $\eta_c$(1S)$\rightarrow K^0_S K^\pm \pi^\mp$, with $K^0_S \rightarrow \pi^+ \pi^-$. The idea is to look for the $X$ states in $\gamma \gamma$ interactions because their observation would add useful information on their decay modes and also on the quantum numbers of the X(3872). A theoretical prediction is that BF($\eta_c$(2S)$\rightarrow \eta_c$(1S)$\pi^+ \pi^-$) = 2.2$\%$\cite{voloshin}.

Our results are reported in  Table 2. No significant signal is observed, but upper limits are established on the existence of the $X$ candidate states in a $\gamma \gamma$ process. We find  that 
BF($\eta_c(2S) \rightarrow \eta_c(1S) \pi^+ \pi^-)<$7.4$\%$ at 90$\%$ confidence level (c.l.), which is consistent with the theoretical prediction.

\begin{table}[!htb]
\caption{UL (90$\%$ c.l.) values obtained from the analysis of $\gamma \gamma \rightarrow X \rightarrow \eta_c \pi^+ \pi^-$ (\babar preliminary).}
\begin{center}
\begin{tabular}{lrccccl}
\hline
\hline
Resonance & M$_X$ (MeV/c$^2$)& $\Gamma_X$ (MeV) & $\Gamma_{\gamma \gamma} \cal B$ (eV) UL\\ \hline
$\chi_{c2}$(1P)   & 3556.20 $\pm$ 0.09  &  1.97 $\pm$ 0.11 & 15.7\\  
$\eta_c$(2S)     & 3638.5 $\pm$ 5.6  &  13.4 $\pm$ 5.6 & 133\\  
X(3872)          & 3871.57 $\pm$ 0.25 & 3.0 $\pm$ 2.1  & 11.1\\ 
X(3915)          & 3915.0 $\pm$ 3.6 &  17.0 $\pm$ 10.4 & 16\\ 
$\chi_{c2}$(2P)   & 3927.2 $\pm$ 2.6  & 24 $\pm$ 6      & 19\\  \hline 
\end{tabular}
\end{center}
\label{table2}
\end{table}
\subsection{Analysis $\gamma \gamma \rightarrow J/\psi \omega$}  
 
$~~~$ An interesting recent result from \babar\cite{antimo} involves the study of the the invariant mass of the $J/\psi \omega$ system produced via $\gamma \gamma$ interactions. After the discovery of the  Y(3940) state by Belle\cite{belle4} in the decay $B^\pm \rightarrow J/\psi \omega K^\pm$, \babar confirmed the existence of a similar resonant state in B decays\cite{babaradd}, but with lower mass and smaller width compared to the Belle results. 

In a re-analysis\cite{babar5} of the \babar data which used the complete $\Upsilon(4S)$ data sample and reduced the lower limit on the $\omega$ signal region to 740 MeV/c$^2$, the precision of the Y(3940) measurements was improved, and evidence for the decay X(3872)$\rightarrow J/\psi \omega $ was reported. This confirmed an earlier unpublished Belle claim for the existence of this decay mode\cite{belleadd}. The latter was based on the behaviour of the invariant $\pi^+ \pi^- \pi^0$ mass distribution near the X(3872) mass, whereas the \babar result is obtained directly from a fit to the $J/\psi \omega$ mass distribution. 

A subsequent paper from Belle~\cite{belle5} reports the observation in $\gamma \gamma \rightarrow J/\psi \omega$ of a state, the X(3915), with  mass and width values similar to those obtained for the Y(3940) in the \babar analysis\cite{babaradd}. 

\begin{figure}[htb]
\centering
\includegraphics[height=2.5in]{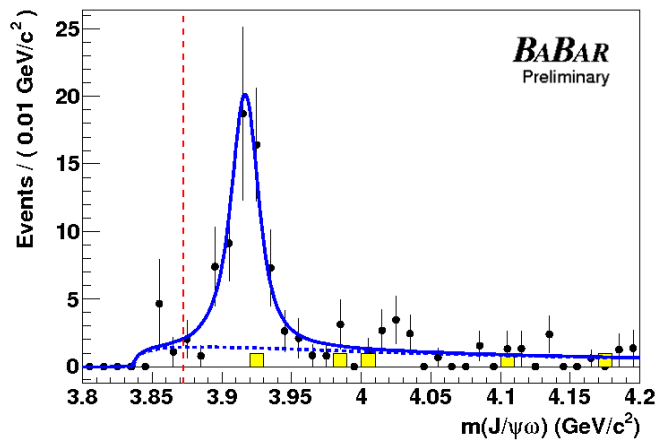}
\caption{Invariant mass of the $J/\psi \omega$ system from the process $\gamma \gamma \rightarrow J/\psi \omega$.}
\label{fig:X3915}
\end{figure}

The preliminary result of the \babar $\gamma \gamma$ analysis is reported in Fig.~\ref{fig:X3915}, where  a resonant state around 3915 MeV/c$^2$ is observed, but no evidence for the X(3872) is found, as in the Belle paper. The peak at 3915 MeV/c$^2$ is observed with 7.6$\sigma$ significance, using the data samples collected from \babar at $\Upsilon$(nS), n=2,3,4, in total 520 fb$^{-1}$. A summary of the fit results is presented in Table 3. A study of the spin-parity assignment of the X(3915) is presently in progress. 
\begin{table}[!htb]
\caption{Fit values obtained for the X(3915) (\babar preliminary).}
\begin{center}
\begin{tabular}{lrccl}
\hline
\hline
X(3915) & \babar& Belle \\ \hline
 M$_X$ (MeV/c$^2$)  & 3919.4 $\pm$ 2.2 $\pm$ 1.6 &  3915 $\pm$ 3$\pm$2 \\  
  $\Gamma_X$ (MeV)  & 13 $\pm$ 6  $\pm$ 3 &  17 $\pm$ 10 $\pm$ 3 \\  
 $\Gamma_{\gamma \gamma} \cal B$ (eV) , J=0      & 52 $\pm$ 10 $\pm$3 & 61 $\pm$ 17 $\pm$8\\ 
$\Gamma_{\gamma \gamma} \cal B$ (eV) , J=2& 10.5 $\pm$ 1.9   $\pm$ 0.6 & 18 $\pm$ 5   $\pm$ 2\\ \hline 
\end{tabular}
\end{center}
\label{table3}
\end{table}
\section{Conclusion}

 $~~~$ We have shown preliminary results from new \babar analyses, in particular from ISR processes and  $\gamma \gamma$ interactions.
We confirm the states Y(4260), Y(4360) and Y(4660) in ISR processes, and also the existence of the X(3915) in the reaction $\gamma \gamma \rightarrow J/\psi \omega$ . The state Y(4008) reported by Belle in $e^+e^- \rightarrow J/\psi \pi^+ \pi^-$ is not confirmed. 

No evidence of the X(3872) was found in $\gamma \gamma \rightarrow J/\psi \omega$. It should not be present if it has J=1, but its absence could be due also to weak coupling to the two-photon initial state. 

Many other analyses are in progress at present, and \babar is still producing very interesting results  4 years after the end of data-taking. 
\\

We are grateful for the excellent luminosity and machine conditions
provided by our colleagues, 
and for the substantial dedicated effort from
the computing organizations that support \babar.
The collaborating institutions wish to thank 
SLAC for its support and kind hospitality. 
This work is supported by
DOE
and NSF (USA),
NSERC (Canada),
CEA and
CNRS-IN2P3
(France),
BMBF and DFG
(Germany),
INFN (Italy),
FOM (The Netherlands),
NFR (Norway),
MIST (Russia),
MEC (Spain), and
STFC (United Kingdom). 
Individuals have received support from the
Marie Curie EIF (European Union) and
the A.~P.~Sloan Foundation.

\end{document}